\magnification\magstep1
\font\ftitle=cmbx10
scaled\magstep2
\def\var{\varepsilon}
\font\smc=cmcsc10

\centerline{\ftitle Dynamics of dissipative two--level systems in the}
\smallskip
\centerline{\ftitle stochastic approximation}\bigskip
\centerline{{\smc L. Accardi$^{1}$, S.V. Kozyrev$^{2}$} and {\smc
I.V. Volovich$^{3}$}}\bigskip\footnote{}
{\hglue-\parindent
$^{1}$Centro Vito Volterra, Universita di Roma Tor Vergata
00133, Italia,
\hfill\break
accardi@volterra.mat.utovrm.it
\hfill\break
$^{2}$ Institute of Chemical Physics, Kossygina St.4,117334,
Moscow, Russia,
\hfill\break
kozyrev@genesis.mian.su
\hfill\break
$^{3}$ Steklov Mathematical Institute of Russian Academy of Sciences,
Gubkin St.8, 117966
Moscow, Russia,volovich@genesis.mi.ras.ru}
\bigskip

\noindent{\bf Abstract}. 
The dynamics of the spin--boson Hamiltonian is
considered in the stochastic approximation. The Hamiltonian describes a
two--level system coupled to an environment and is widely used in
physics, chemistry and the theory of quantum measurement.

We demonstrate that the method of the stochastic
approximation which is a general method of consideration of dynamics of
an arbitrary system interacting with environment is powerful enough to
reproduce qualitatively striking results by Leggett at al. found
earlier for this model. The result include an exact expression of the
dynamics in terms of the spectral density and show an appearance of two
most interesting regimes for the system, i.e. pure oscillating and pure
damping ones. Correlators describing environment are also
computed.\bigskip\medskip

\beginsection{1. Introduction}

\bigskip
The so--called spin--boson Hamiltonian is widely used in physics and
chemistry. In its simplest version it describes a
dynamical model of a two--level system coupled to an
environment. One of the basic ideas is that the environment induces
dissipative effects, but as we shall see, the picture is much richer.
Examples include
the motion of defects in some crystalline solids, the motion of the
magnetic flux trapped in an $rf$ SQUID ring, some chemical reactions,
some approaches to the
theory of quantum measurement and many other quoted in the survey paper
[1] to which the present work is inspired.
The ``spin--boson'' Hamiltonian considered in the present paper is the
same considered in [1], i.e.
$$H_\lambda=-{1\over2}\,\Delta\sigma_x+{1\over2}\,\var\sigma_z+\int
dk\omega(k)a^+(k)a(k)+\lambda\sigma_z(A(g^*)+A^+(g))\eqno(1.1)$$
where $\sigma_x$ and $\sigma_z$ are Pauli matrixes, $\var$ and $\Delta$ are
real parameters interpreted respectively as the
energy difference of the states localized in the two wells in
absence of tunneling and as the matrix element for
tunneling between the wells. We set $\Delta>0$ and deno
te
$$A^+(g)=\int a^+(k)g(k)dk\ ,\qquad A(g^*)=\int a(k)g^*(k)dk$$
where $a(k)$, and $a^+(k)$ are bosonic annihilation and creation
operators
$$[a(k),a^+(k')]=\delta(k-k')$$
which describe the environment.

We denote $\omega(k)$ the one--particle energy of the environment and assume
$\omega(k)\geq0$. The function $g(k)$ is a form factor describing the
interaction of the
system with the environment, $\lambda$ is the coupling constant.
It is well known that, in times of order
$t/\lambda^2$, the interaction produces effects of order $t$. Thus
$\lambda$ provides a natural time scale for the observable effects of
the interaction system--environment.

In the paper [1] it was found a very rich behavior of the dynamics of the
Hamiltonian (1)
ranging from undamped oscillations, to exponential relaxation, to
power--law types of behavior and to total localization. Leggett et al. [1]
found the remarkable result that main qualitative features of the
system dynamics can be described in terms of the
temperature (i.e. the initial state of the environment) and of the
behavior, for low frequencies $\omega$, of the spectral function
$$J(\omega):=\int dk|g(k)|^2\delta(\omega(k)-\omega)\eqno(1.2)$$
The goal of this paper is to investigate the dynamics of the Hamiltonian
(1) in the so called {\it stochastic approximation\/}. The overall qualitative
picture emerging from this approach is similar to the one described in
[1] and in some cases also the quantitative agreement is good (cf.
Section [4]).\bigskip\medskip

\beginsection{2. The stochastic approximation}

\bigskip
The basic idea of the stochastic approximation is the following.
If one has a Hamiltonian of the form
$$H_\lambda=H_0+\lambda V\eqno(2.1)$$
then, by definition the stochastic limit of the evolution operator
$$U^{(\lambda)}(t)=e^{itH_0}e^{-itH_\lambda}\eqno(2.2)$$
is the following limit (when it exists in the sense specified by (2.11)
and (2.12) below):
$$U(t)=\lim_{\lambda\to\infty}U^{(\lambda)}\left({t\over\lambda^2}\right)
\eqno(2.3)$$
Notice that, on the right hand
side of (2.3), there is not
$U^{(\lambda)}_t$ but its rescaled version $U^{(\lambda)}(t/\lambda^2)$.
Thus the
limiting evolution operator $U(t)$ (2.3) describes the
behavior of the model in the time scale described in the introduction.
The stochastic
approximation is a natural generalization of the Friedrichs-van Hove limit
which uses the same time rescaling but allows only to compute vacuum
expectation values of the form $\langle U^{(\lambda)}_{t/\lambda^2}
\Lambda U^{(\lambda)*}_{t/\lambda^2}\rangle$ for particular classes of
observables $\Lambda$. This leads to {\bf irreversible evolution} and to the
corresponding {\bf master equation}. On the contrary, the stochastic
approximation leads to {\bf reversible, unitary evolution} and to the
corresponding {\bf quantum stochastic differential equation} from which
the master equation are deduced by a now standard procedure which
consists in {\it integrating away\/} the environment degrees of freedom.

The stochastic approximation to the original dynamics (2.2) consists in
the
computation of the limit (2.3) in the sense of matrix elements over some
states $\psi_\lambda$ (called ``collective states'') which  themself
depend on the parameter $\lambda$ in a singular way. The fact that one
cannot expect the limit (2.3) to exist for arbitrary states, but only
for a carefully chosen class of states was already pointed out in the
classical paper of van Hove [2]. The effective determination of this
class of states was obtained in the paper [3].

The stochastic approximation could also be considered as a new kind of
semiclassical approximation in the sense that it studies the {\it
fluctuations\/} around the classical solution and not the approximation
to it. This interpretation however shall not be discussed here (cf. [4]).

One of the important features of the stochastic method is its {\it
universality\/}. The restriction to Pauli matrixes in (1.1) is
unnecessary: the theory is applicable
whenever the evolution operator $U^{(\lambda)}(t)$ (2.2)
satisfies the equation
$${dU^{(
\lambda)}(t)\over dt}=-i\lambda V(t)U^{(\lambda)}(t)\eqno(2.4)$$
where $V(t)=e^{itH_0}V e^{-itH_0}$ has the form
$$V(t)=\sum_\alpha(D^+_\alpha\otimes A_\alpha(t)+D_\alpha\otimes
A^+_\alpha(t))\eqno(2.5)$$
and the $D_\alpha$ are operators describing the system.
The rescaled evolution operator $U^{(\lambda)}\left({t\over\lambda^2}
\right)$, associated to (2.5), satisfies the equation
$${dU^{(\lambda)}\left({t\over\lambda^2}\right)\over dt}\,=-i\sum_\alpha
\left(D^+_\alpha\otimes{1\over\lambda}\,A_\alpha\left({t\over\lambda^2}\,
\right)+D_\alpha\otimes{1\over\lambda}\,A^+_\alpha\left({t\over
\lambda^2}\right)\right)
U^{(\lambda)}(t/\lambda^2)\eqno(2.7)$$
In the spin--boson Hamiltonian (1.1) the $D_\alpha$ are Pauli matrixes
(cf. formulae (3.10b), (3.12b) (3.8)) and
$$A_\alpha(t)=\int a(k)e^{-it\omega_\alpha(k)}g^*(k)dk\eqno(2.6a)$$
where the functions $\omega_\alpha(k)$ have the form
$$\omega_\alpha(k)=\omega(k)-\omega_\alpha\ ;
\quad\alpha=1,2,3\eqno(2.6b)$$
here $\omega(k)$ is as in (1.1) and the $\omega_\alpha$ are
characteristic frequencies given by formula (3.12a).

>From (2.7) it is clear that to have a nontrivial limit for
$U^{(\lambda)}(t/\lambda^2)$, the limit
$$\lim_{\lambda\to0}{1\over\lambda}\,A_\alpha\left({t\over\lambda^2}\,\right)=
b_\alpha(t)\eqno(2.8)$$
should exist. It can be proved (cf. [4]) that
the limit (2.8) exists for ``good'' functions $\omega_\alpha(k)$
and $g(k)$ in the sense that
$$\lim_{\lambda\to0}\langle{1\over\lambda}\,A^{\var_1}_{\alpha_1}
\left({t_1\over\lambda^2}\right)\dots{1\over\lambda}\,A^{\var_n}_{\alpha_n}
\left({t_n\over\lambda^2}\right)\rangle=\langle b^{\var_1}_{\alpha_1}
(t_1)\dots b^{\var_n}_{\alpha n}(t_n)\rangle\eqno(2.9)$$
where the indices $\var_i$ label the creators $(\varepsilon=0)$
and the annihilators $(\varepsilon=1)$; the brackets in (2.9)
denote mean values over the Fock vacuum or a temperature state; and
for each $\alpha$, $b_\alpha(t)$ is the Fock Boson quantum field
described by (2.14), (2.15) below. In the literature
$\delta$--correlated (in time) quantum
fields are often called {\bf
quantum noises}. In the present paper only quantum white noises shall
appear, but it is important to keep in mind that many other
possibilities can arise from different physical models.

>From (2.7) one has in the limit $\lambda\to0$:
$${dU(t)\over dt}\,=-i\sum_\alpha(D^+_\alpha\otimes b_\alpha(t)
+D_\alpha\otimes b^+_\alpha(t))U(t)\eqno(2.10)$$
The limit (2.3) means that:
$$\lim_{\lambda\to0}\langle\Psi_\lambda,U^{(\lambda)}
\left({t\over\lambda^2}\right)\Psi'_\lambda\rangle=\langle\psi,
U(t)\psi'\rangle\eqno(2.11)$$
where the collective vectors $\Psi_\lambda$ are defined by
$$\Psi_\lambda={1\over\lambda}\,A^+_{\alpha_1}\left({t_1\over\lambda^2}\right)
\dots{1\over\lambda}\,A^+_{\alpha_n}\left({t_n\over\lambda^2}\right)
\Psi^{(0)}\ ,\eqno(2.12)$$
and converge to the corresponding $n$--particle vectors in the noise
space, given by:
$$\psi=b^+_{\alpha_1}(t_1)\dots
b^+_{\alpha_n}(t_n)\psi^{(0)}\eqno(2.13)$$
$\Psi^{(0)}$ and $\psi^{(0)}$ are the vacuum vectors in the corresponding
Fock spaces. If $\omega_\alpha\not=\omega_\beta$ for $\alpha\not=\beta$
(as it is the case for the Hamiltonian (1.1)) then $b_\alpha$, $b^+_\alpha$
satisfy the following commutation relations
$$[b_\alpha(t),b^+_{\alpha'}(t')]=\delta_{\alpha\alpha'}J_\alpha
\delta(t-t')\eqno(2.14)$$
where $J_\alpha$ is the spectral function (1.2)
$$J_\alpha=2\pi\int dk|g(k)|^2\delta(\omega_\alpha(k))\eqno(2.15)$$
Thus, as announced in the introduction, in the stochastic limit the
spectral function emerges naturally as the covariance of the quantum
noise. Some care is needed in the interpretation of equation (2.10)
because, as it is clear from (2.14) the $b_\alpha(t)$ are not {\it bona
fide\/} operators but only operator valued distribution. In order to
give a meaning to equation (2.10) (more precisely to its matrix elements
in the $n$--particle or coherent vectors), we
rewrite (2.10) in normal form by bringing $b_\alpha(t)$ to the right of
$U(t)$. This gives rise to a commutator which can be explicitly computed.
 The result is:

$${dU(t)\over dt}\,=-i\sum_\alpha(D^+_\alpha U(t)b_\alpha(t)+D_\alpha
b^+_\alpha(t)U(t)-i\gamma_\alpha D^+_\alpha D_\alpha U(t)\eqno(2.16)$$
where $\gamma_\alpha$ are complex numbers given explicitly by:
$$\gamma_\alpha=\int^0_{-\infty}d\tau\int dke^{i\tau\omega_\alpha(k)}
|g(k)|^2\eqno(2.17)$$
The connection between the constants $\gamma_\alpha$ in
the last term in (2.16) ( the Ito
correction term), and the spectral function (2.15) is obtained by
exchanging the $d\tau$--and the $dk$--integral in (2.17) and using the
known formula
$$\int^0_{-\infty}e^{it\omega}dt=\pi\delta(\omega)-iP.P.{1\over\omega}$$
where $P.P.$ denotes the principal part integral. This shows that {\it
the spectral functions are the real parts of the constants
$\gamma_\alpha$, emerging in the Ito correction term\/}. This connection
is the prototype of the {\it dispersion relations\/} widely used in
quantum physics since its origins.
Since the $\gamma_\alpha$ are complex, equation (2.16) {\it looks
like\/} an equation driven by a non self--adjoint Hamiltonian. However
this is only an apparent phenomenon due to the normal order. The true
Hamiltonian (2.10), although singular, is formally self--adjoint and
this gives an intuitive explanation of the unitarity of the solution of
(2.16) or, equivalently, of (2.10).

The relations (2.14)--(2.17) define the stochastic approximation to the system
(2.4), (2.5). The term {\bf stochastic} is justified by the fact that
the distribution equation (2.16) which has a weak meaning in the
$n$--particle vectors, can be interpreted as a quantum stochastic
differential equation (and, in fact, it is in this form that this
equation was first derived [3]). The operators
$b_\alpha(t)$, $b^+_\alpha(t)$ are called a {\bf quantum white noise}
and the additional term in (2.16), arising in (2.16) from normal order is
called the {\it drift\/} or the {\it Ito correction term\/}.
More precisely, in quantum probability one usually writes
(2.16) in the form
$$dU(t)=-i\sum_\alpha(D^+_\alpha dB_\alpha(t)+D_\alpha dB^+_\alpha
(t)-i
\gamma_\alpha D^+_\alpha D_\alpha dt)U(t)\eqno(2.18)$$
where
$$dB_\alpha(t)=\int^{t+dt}_tb_\alpha(\tau)d\tau$$
are called stochastic differentials and satisfy the Ito table:
$$dB_tdB^+_t=2\gamma dt\ ;\quad dtdB^{+}_t=dB_tdB_t=dB^+_tdB^+_t=
dB^+_tdB_t=0\eqno(2.19)$$
The proof of the Ito table (2.19), as well as its rigorous meaning, was
first established in [5]. This has been subsequently applied to several
models in quantum optics by [3] and [4]. Using it, the unitarity of the
solution of (2.18) is easily established.

The advantage of equation (2.16) over the original one (2.4) is that it
is in some sense completely integrable and one
can easily read the physics from it.
For example for the vacuum expectation value one has the equation:
$${d\langle U(t)\rangle\over dt}\,=-\sum_\alpha\gamma_\alpha D^+_\alpha
D_\alpha\langle U(t)\rangle$$
which gives the damped oscillatory regime (the $\gamma_\alpha$ are
complex number):
$$\langle U(t)\rangle=e^{-\Gamma t}\ ,\quad\Gamma=\sum_\alpha\gamma_\alpha
D^+_\alpha D_\alpha$$
In the following we shall apply this method to the
Hamiltonian (1.1) and, in Section (4), we shall compare our result with
those of [1], [6].\bigskip\medskip

\beginsection{3. The stochastic approximation for the ``spin--boson''
system}

\bigskip
In order to apply the stochastic approximation to the Hamiltonian (1.1),
we write (1.1) in the form (2.1) where
$$H_0=H_S+H_R\eqno(3.1)$$
The system Hamiltonian $H_S$ is
$$H_S=-{1\over2}\,\Delta\sigma_x+{1\over2}\,\var\sigma_z\eqno(3.2)$$
and the reservoir Hamiltonian $H_R$ is
$$H_R=\int dk\omega(k)a^+(k)a(k)\eqno(3.3)$$
The evolution operator $U^{(\lambda)}(t)$ satisfies equation (2.4) where
$$V(t)=\sigma_z(t)(A(e^{-it\omega}g^*)+A^+(e^{it\omega}g))\eqno(3.4)$$
and
$$\sigma_z(t)=e^{itH_S}\sigma_ze^{-itH_S}\eqno(3.5)$$
To bring (3.4) to the form (2.5) let us compute (3.5). The eigenvalues
of the Hamiltonian (3.2) are
$$H_S|e_\pm>=\lambda_\pm|e_\pm>\eqno(3.6)$$
where
$$\lambda_\pm=\pm{1\over2}\,\Delta\nu\eqno(3.7)$$
$$|e_\pm>={1\over\sqrt{1+\mu^2_\mp}}\pmatrix{
1\cr
\mu_\mp\cr}\eqno(3.8)$$
and
$$\mu_\pm={\var\over\Delta}\,\pm \nu\ ,\quad \nu=\sqrt{1+\left({\var
\over\Delta}\right)^2}\eqno(3.9)$$
Notice, for future use, that:
$$\langle e_\pm|\sigma_z|e_\pm\rangle={1-\mu^2_\mp\over1+\mu^2_\mp}\quad;
\qquad
\langle e_+|\sigma_z|e_-\rangle=\langle e_-|\sigma_z|e_+\rangle=1/\nu$$
Therefore
$$\sigma_z(t)={1-\mu^2_-\over1+\mu^2_-}\,DD^++{1-\mu^2_+\over1+\mu^2_+}\,
D^+D+\nu^{-1}e^{it \nu\Delta}\,D+ \nu^{-1}e^{-it\nu\Delta} D^+
\eqno(3.10a)$$
where
$$D=|e_+><e_-|\eqno(3.10b)$$
The interaction Hamiltonian (3.4) can now be written in the form (2.5):
$$V(t)=\sum^3_{\alpha=1}(D^+_\alpha\otimes A(e^{-it\omega_\alpha}
g^*)+h.c.)\eqno(3.11)$$
where the three spectral frequencies correspond respectively to the
down, zero, and up transitions of the 2--level system, i.e.
$$\omega_1(k)=\omega(k)-\nu\Delta\quad ;\qquad
\omega_2(k)=\omega(k)\quad ;\qquad\omega_3(k)
=\omega(k)+\nu\Delta\eqno(3.12a)$$
$$D_1=\nu^{-1}D^+\quad ;
\qquad D_2={1-\mu^2_-\over1+\mu^2_-}\ DD^++{1-\mu^2_+\over1+
\mu^2_+}\, D^+D\quad ;\qquad D_3=\nu^{-1}D^+\eqno(3.12b)$$
The corresponding limiting evolution equation therefore has the form (2.16).
It is important to note however that the constants (2.15) for $\alpha=2,3$
vanish, i.e.
$$J_2=J_3=0\eqno(3.13)$$
We shall see that the purely oscillatory regime, first discovered by
Leggett et al. [1] corresponds to the case when also $J_1$ vanishes. In
this sense it can be interpreted as an {\it off--resonance\/} regime. In
this regime a strange (from the point of view of stochastic theory) new
phenomenon take place: in $t/\lambda^2$--limit the environment
disappears (i.e. the limit on the right hand side of (2.8) is zero,
corresponding to a quantum white noise of zero variance).  However a
remnant of the interaction remains because, after the limit, the system
evolves with a new hamiltonian, equal to the old one plus a shift term
depending on the interaction and on the initial state of the field. This
is a kind of {\it Cheshire Cat effect\/}.

(3.13) implies that the operator
s $b_2$ and $b_3$ should be absent in (2.17).
However the constants $\gamma_2$ and $\gamma_3$ as well as $\gamma_1$ do
contribute to (2.16). We denote $b_1(t)$ by $b(t)$. Thus the operators
$b(t)$, $b^+(t)$ satisfy
$$[b(t),b^+(t')]=\gamma\delta(t-t')\eqno(3.17)$$
with $\gamma$ given by (3.15) below and $\nu$ (in $\gamma$) given by
(3.9). The limiting evolution equation can then be written:
$${dU(t)\over dt}\,=Db^+(t)U(t)-D^+U(t)b(t)-(\gamma+i\sigma)
D^+DU(t)-i\varphi U(t)\eqno(3.14)$$
where
$$\gamma=\nu^{-2}\pi J(\nu\Delta)\ ,\eqno(3.15)$$
$$\sigma=\nu^{-2}(I(-\nu\Delta)-I(\nu\Delta))+\left(\left({1-\mu^2_-\over1+
\mu^2_-}\right)^2-\left({1-\mu^2_+\over1+\mu^2_+}\right)^2\right)I(0)\ ,$$
$$\varphi=\nu^{-2}I(-\nu\Delta)+\left({1-\mu^2_-\over1+\mu^2_-}
\right)^2I(0)$$
and we denote
$$J(\omega)=\int dk|g(k)|^2\delta(\omega(k)-\omega)\quad;\qquad
I(\omega)=P.P.\int^\infty_0{d\omega' J(\omega')\over\omega'-\omega}
\eqno(3.16)$$
where $P.P.$ means the principal part of the integral.

In the notations of quantum stochastic equations (3.14) reads
$$dU(t)=(DdB^+_t-D^+dB_t-(\gamma+i\sigma)D^+D-i\varphi)U(t)\eqno(3.18)$$
Notice that all parameters $\gamma$, $\sigma$ and $\varphi$ in the
evolution equation (3.14) are expressed in terms of the spectral density
$J(\omega)$ (3.16) and parameters $\Delta$ and $\varepsilon$ of the
original Hamiltonian (1.1).\bigskip\medskip

\beginsection{4. Analysis of the stochastic approximation. Zero
temperature}

\bigskip
Let us discuss now in more detail the implications of the results of the
previous sections
for the ``spin--boson'' Hamiltonian. All the information
about the model is encoded into the constants $\gamma$, $\sigma$ and
$\varphi$ and these constants are expressed in terms
of the spectral density $J(\omega)$ (3.16)
depending on the parameters of the
Hamiltonian ($\var$ and $\Delta$) and the temperature (not yet
introduced up to now).
Thus the method of stochastic approximation confirms the conclusion of
Leggett at al. [1] that the long--time behavior of the model is
expressed i
n terms of the spectral density $J(\omega)$.

Now let us discuss the dynamics of the system in the stochastic
approximation. We are interested in a pure damping or pure oscillating
behavior.

For the vacuum expectation value we have
$$\langle U(t)\rangle=e^{-i\varphi t}+e^{-i\varphi t}(e^{-(\gamma+i
\sigma)t}-1)D^+D\eqno(4.1)$$
and taking trace over the spin variables one gets (since $TrD^+D=1$ ):
$$\langle tr U(t)\rangle=e^{-[\gamma+i(\sigma+\varphi)]t}\eqno(4.2)$$
Since $\gamma$, $\sigma$, and $\varphi$ are real (cf. (3.14)--(3.16)),
one has a purely oscillating behavior (4.2) if and only if there is no
damping, i.e.
$$\gamma=0\eqno(4.4)$$
However one cannot have a vanishing of oscillations, because the
quantity
$$\sigma+\varphi=\nu^{-2}I(-\nu\Delta)+\left({1-\mu^2\over1+\mu^2}
\right)^2I(0)>0\eqno(4.3)$$
is strictly positive for positive $J(\omega)$ (because $\nu$,
$\Delta>0$, cf. (3.9)) and $I(\omega)$ is given by (3.16).

The stochastic approximation to the vacuum expectation value of the
Heisenberg evolution of $\sigma_z$ is given by
$$P(t)=\langle U^*(t)\sigma_z(t)U(t)\rangle\eqno(4.4)$$
>From equation (3.14) one gets the Langevin equation for $P(t)$ which
solution is
$$P(t)=\nu^{-1}e^{-\gamma t}(D^+e^{i(\sigma-\nu\Delta)t}+De^{-i
(\sigma-\nu\Delta)t})+$$
$$+D^+D\left({1-\mu^2_+\over1+\mu^2_+}\,-{1-\mu^2_-\over1+\mu^2_-}\right)
e^{-2\gamma t}+{1-\mu^2_-\over1+\mu^2_-}\eqno(4.5)$$
Let us discuss separately the simplest case $\varepsilon=0$.\bigskip

\noindent{\it The case $\varepsilon=0$, zero temperature}\medskip

In this case one has
$$P(t)=e^{-\gamma t}(D^+e^{i(\sigma-\Delta)t}+De^{-i(\sigma-\Delta)t})
\eqno(4.6)$$
where $\gamma$, $\sigma$, and $I(\omega)$ are now
$$\gamma=\pi J(\Delta)\quad ;\qquad\sigma=I(-\Delta)-I(\Delta)\quad ;\qquad
I(\omega)=P.P.\int{d\omega' J(\omega')\over\omega'-\omega}$$
Two interesting regimes can now appear:
\item{(i)} {\it No oscillations\/}. In this case
$$\sigma-\Delta=0\eqno(4.7)$$
Equation (4.7) is equivalent to the integral equation
$$\int{dxJ(x)\over x+\Delta}\,-P.P.\
int{dx J(x)\over x-\Delta}\,
=\Delta\eqno(4.8)$$
If equation (4.8) is satisfied then we have pure damping:
$$P(t)=e^{-\gamma t}(D^++D)\eqno(4.9)$$

We will discuss solutions of eq. (4.8) later.\bigskip

Another regime is
\item{(ii)} {\it Pure oscillations\/}. This regime is defined by the
condition
$$\gamma=\pi J(\Delta)=0\eqno(4.10)$$
Notice that, because of (3.16) this condition defines an {\bf
off--resonance condition}.

If equation (4.10) satisfied, then
$$P(t)=D^+e^{i(\sigma-\Delta)t}+De^{-i(\sigma-\Delta)t}\eqno(4.11)$$
where
$$\sigma-\Delta=\int{dxJ(x)\over x+\Delta}\,-P.P.\int{dxJ(x)
\over x-\Delta}\,-\Delta$$
This case of pure oscillations is very interesting. If there is a
damping then after a rather short time $P(t)$ becomes a small quantity
which is difficult to observe. The case of permanent oscillations looks
more promising for observations. This regime is of primary interest in
the context of the so--called macroscopic quantum coherence phenomenon
[7].

The purely oscillatory regime was discovered in [1] but the region of
parameters there is different from ours. To get pure oscillations we
need the only off--resonance condition (4.10), i.e. in terms of the spectral
density what we need is:
$$J(\Delta)=\int dk|g(k)|^2\delta(\omega(k)-\Delta)=0$$
The difference with [1] can be attributed to the different boundary
conditions on correlators.

Let us present our results on the computation of the correlator
$$C(t)={1\over2}\,\langle\{U^*_t\sigma_zU_t,\sigma_z\}\rangle=
{1\over2}\,\{P(t),P(0)\}$$
We have
$$C(t)={1\over2}\,e^{-(\gamma+i\sigma+i\nu\Delta)t}\left(\nu^{-2}
+\nu^{-1}D\biggl({1-\mu^2_-\over1+\mu^2_-}\,+{1-\mu^2_+\over1+
\mu^2_+}\right)\eqno(4.12)$$
$$+h.c.+{1-\mu^2_-\over1+\mu^2_-}\,(\nu^{-1}(D+D^+)+{1-\mu^2_-
\over1+\mu^2_-}\,DD^++{1-\mu^2_+\over1+\mu^2_+}\,D^+D\biggr)+$$
$$+e^{-2\gamma}\,\left({1-\mu^2_+\over1+\mu^2_+}\,-{1-\mu^2_-
\over1+\mu^2_-}\,\right)(\nu^{-1}(D+D^+)+
2D^+D{1-\mu^2_+\over1+\mu^2_+}\biggr)$$
The trace of $C(t)$ is
$$trC(t)=2\nu^{-2}e^{-\gamma t}\,\cos(\sigma+\nu\Delta)t+
2e^{-2\gamma t}\left({1-\mu^2_+\over1+\mu^2_+}\,-
{1-\mu^2_-\over1+\mu^2_-}\right)\left({1-\mu^2_+\over1+\mu^2_+}\right)$$
The qualitative behavior of $C(t)$ is such as for $P(t)$.\bigskip

\noindent{\it Non--zero temperature}\medskip

For a non--zero temperature we get a stochastic evolution equation of
the same form as before (3.14) only with new constants $\gamma$, $\sigma$
and $\varphi$. More precisely:
$$\gamma=\nu^{-2}\pi(J_+(\nu\Delta)+J_-(\nu\Delta))$$
$$\sigma=\left[\left({1-\mu^2_+\over1+\mu^2_+}\right)^2-
\left({1-\mu^2_-\over1+\mu^2_-}\right)^2\right](I_+(0)+I_-(0))+$$
$$+\nu^{-2}(I_+(-\nu\Delta)-I_+(\nu\Delta)+I_-(-\nu\Delta)-I_-
(\nu\Delta))$$
where spectral densities are
$$J_+(\omega)={J(\omega)\over1-e^{-\beta\omega}}\qquad;\qquad
J_-(\omega)={J(\omega)e^{-\beta\omega}\over1-e^{-\beta\omega}}$$
Here $J(\omega)$ is the spectral density (3.16)
and $\beta$ is the inverse temperature.

The functions $I_\pm(\omega)$ are defined by
$$I_\pm(\omega)= P.P.\int{d\omega'J_\pm(\omega')\over\omega'-
\omega}$$
One has the same as for the zero--temperature expressions (4.15)
and (4.12) for
$P(t)$ and $C(t)$ but now with new constants $\gamma$ and $\sigma$
depending on temperature:
$$\gamma=\nu^{-2}\pi J(\nu\Delta)\coth{\beta\nu\Delta\over2}$$\bigskip\medskip

\beginsection{Conclusion}

\bigskip
 To conclude the following main result are obtained:

The theoretical role of the spectral function is explained
through its emergence from a canonical limit procedure. Moreover this
function is shown to be real part of a complex function whose imaginary
part defines an energy shift in the system Hamiltonian. When the
environment free energy depends only on the modulus of momentum
($\omega(k)=\omega(|k|)$ in (1.1)) the real and imaginary part of this
function are related by a Hilbert transform, thus making a bridge with
the standard {\it dispersion relations\/} (cf. Section (2.)).

In the stochastic approach not only the Heisenberg equation
of the system observables is controlled, but also the environment
evolution. It is
 shown that the environment converges to a {\it quantum
noise\/} (a master field, in the particle physicists terminology). This
gives a theoretical (i.e. based on a microscopic Hamiltonian
description) foundation to the use of classical of quantum noises widely
used in several contemporary approaches to quantum measurement theory
[1], [7].
We can compute the limit
matrix elements of Heisenberg evolution for arbitrary $n$--particle or
coherent vectors. The vacuum matrix elements give rise to the master
equation. The control of the other matrix elements is a new feature of
the stochastic approach.

The purely oscillatory regime, discovered by Leggett et al. [1]
is related here to a {\it Cheshire Cat effect\/} in which the
environment variables vanish in the limit but the interaction leaves a
track in the system behavior in the form of an operator shift in the
system Hamiltonian (cf. Section (4.)).\bigskip\medskip

\centerline{\smc Acknowledgments}\bigskip
S.K.and I.V. are grateful to the V.Volterra Center of
the Rome University Tor Vergata where this work was done
for the hospitality. S.K. is supported in part
by the grant RFFI N 95-03-08838. I.V. is supported in part
by the grant RFFI N 960100312.\bigskip\bigskip

\centerline{\smc Bibliography}\bigskip

\item{1} A.J.Leggett, S.Chakravarty, A.T.Dorsey, M.P.A.Fisher,
A.Garg and W.Zwerger,
1987, Rev.Mod.Phys.59, N1,pp.1-85.

\item{2} L.Van Hove, 1955, Physica,21,617.

\item{3} L.Accardi, A.Frigerio and Y.G.Lu, 1990,
Comm.Math.Phys.131,537.

\item{4} L.Accardi, Y.G.Lu and I.Volovich, Quantum
Theory and Its Stochastic Limit, 1997, Oxford University
Press (to be published)

\item{5} R.L.Hudson and K.R.Parthasaraty, 1984, Comm.Math.Phys.
93,301.

\item{6} A.O. Caldeira and A.J. Leggett, 1981, Phys. Rev. Lett. 46, 211.

\item{7} A.J.Leggett and Anupam Garg, 1985, Phys. Rev. Lett. 54, 857.

\bye